\documentstyle[12pt,twoside,cmp209_RM_2]{article}
\RequirePackage{cite}
\RequirePackage[english]{babel}
\font\frak=eufm10  scaled \magstep1 
\font\sfrak=eufm7  scaled \magstep1 
\font\bfrak=eufb10 scaled \magstep1 
\font\sssm=cmss8   scaled \magstep1 
\font\ssb=cmssbx10 scaled \magstep1 
\chardef\atcode=\catcode`\@\catcode`\@=11
\expandafter\let\expandafter\protect\csname protect\endcsname
\def\allowhyphens{\penalty\@M \hskip\z@skip}
\def\set@low@box#1{\setbox\tw@\hbox{,}\setbox\z@\hbox{#1}%
  \setbox\z@\hbox{\dimen@\ht\z@ \advance\dimen@ -\ht\tw@
      \lower\dimen@\box\z@}%
  \ht\z@\ht\tw@ \dp\z@\dp\tw@}
\def\save@sf@q#1{{\ifhmode \edef\@SF{\spacefactor\the\spacefactor}\else
  \let\@SF\empty \fi \leavevmode #1\@SF}}
\def\@glqq{\save@sf@q{\set@low@box{''\/}\box\z@\kern-.04em\allowhyphens}}
\def\ulq{\protect\@glqq}
\def\@grqq{\save@sf@q{\kern-.07em``\kern.07em}}
\def\urq{\protect\@grqq}
\catcode`\@=\atcode
\def\NO{{\setbox1=\hbox{N\/}\dimen1=\ht1
\setbox0=\hbox{\the\scriptfont2\char"0E}
\setbox4=\hbox to\wd0 {\hss\vrule width.45\wd0 height.1\ht0 depth0pt \hss}%
\setbox2=\vbox{\baselineskip=0pt
        \lineskip=.2\ht0
        \box0
        \box4}%
\advance\dimen1 by-.8\ht2
\setbox3=\hbox{\unhbox1\raise\dimen1\box2\kern.1em}\unhbox3}}
\newbox\stacka\setbox\stacka=\hbox{$\stackrel{\scriptscriptstyle\downarrow}{\hbox{\ssb A}}$}
\edef\stack{\vrule width 0pt depth 0pt height \ht\stacka}
\def\sa{\copy\stacka}
\def\A{\hbox{\ssb A}}
\def\B{\hbox{\ssb B}}
\def\c{\hbox{\ssb c}}
\def\bs#1{\mbox{\ssb #1}}
\def\fr#1{\mbox{\frak #1}}
\def\sr#1{\mbox{\sfrak #1}}
\def\br#1{\mbox{\bfrak #1}}
\def\eisf#1{\hbox{\sssm#1}}
\def\bcdot{\mbox{\boldmath$\cdot$}}
\def\bpartial{\mbox{\boldmath$\partial$}}
\def\bmat#1{\mbox{\boldmath$#1$}}
\def\ubm#1{\mbox{\unboldmath$#1$}}
\def\bkey#1{\hbox{\boldmath$#1$}}
\newbox\abox\setbox\abox=\hbox{\ssb A}
\newbox\Elocity\setbox\Elocity=\hbox{\ssb v}
\newbox\PrimE\setbox\PrimE=\hbox{{\ssb v}$^{\prime}$}
\newbox\elocity\setbox\elocity=\hbox{\sssm v}
\newbox\Prime\setbox\Prime=\hbox{{\sssm v}$^{\scriptscriptstyle\prime}$}
\catcode`A=13\def A{\copy\abox}\catcode65=11
\catcode`V=13\def V{\copy\Elocity}\catcode86=11
\catcode`W=13\def W{\copy\PrimE}\catcode87=11
\catcode`v=13\def v{\copy\elocity}\catcode118=11
\catcode`w=13\def w{\copy\Prime}\catcode119=11
\catcode`"=13\let "=\wedge\catcode34=12
\catcode`M=13\def M{\bf\Omega}\catcode77=11
\def\b{\begingroup\catcode`A=13 \catcode`V=13 \catcode`W=13 \catcode`M=13
\catcode`v=13 \catcode`w=13\catcode`"=13}
\def\e{\endgroup}
\let\lable=\label
\newdimen\colsep\colsep=\arraycolsep
\newcommand{\DS}{\displaystyle}
\newcommand{\SS}{\scriptscriptstyle}
\newtheorem{prop}{Proposition}
\title[THIRD-ORDER CLASSICAL SPIN DYNAMICS IN PLANE]%
     {THIRD-ORDER RELATIVISTIC
     DYNAMICS: CLASSICAL SPINNING PARTICLE TRAVELLING IN A PLANE}
\author{Roman Matsyuk}
\address{Institute for Applied Problems in Mechanics and Mathematics,
        \\ National Academy of Sciences.
        \\ 15~Dudayev~Str., 290005 L\kern-1pt'viv, Ukraine
        \\ matsyuk@lms.lviv.ua}
\date{Received March 12, 1998}
\begin{document}
\maketitle

\begin{abstract}
Mathisson's `new mechanics' of a relativistic spinning particle is shown to
follow, in the case of planar motion, from only general requirements of
relativistic invariance and of the dependence on third order derivatives
along with the `variationality' feature. The hamiltonian counterpart
ultimately recovers the Dixon equations for this case with
the Pirani supplementary condition.
\pacs 03.20.+i 02.40.+m
\end{abstract}

\section{INTRODUCTION}
Following the updating tendencies in the formal theory of variational
calculus promoted by the development of the intrinsic differential geometry
as well as global analysis, many authors have revisited the Ostroghrads'kyj
mechanics with higher derivatives. The subject develops continuously and,
surprisingly enough, models of physical meaning breed (see \cite{G} and
references in \cite{L}).

In this paper we consider a model third-order dynamics of a classical
particle which, although restricted to the unrealistic three-dimensional flat
space-time, provides an instructive example of how new hamiltonian systems
of physical meaningfulness may arise from higher-order variational calculus.
This example admits a comprehensive solution of the involved variational
inverse problem for invariant third-order equation of motion. It turns out
that the equation thus obtained may be interpreted as yet another description
of a planar motion of the classical spinning particle in special relativity.
Taking into account that the general-relativistic equation of motion of the
gravitational dipole particle admits, among others, also a solution of only
two degrees of freedom (see \cite{P1,P2}), we hope that the results of
present investigation may contribute to the future lagrangian and hamiltonian
formulation of the general Mathisson equation \cite{M} in the realistic
four-dimensional curved space-time:
\begin{eqnarray}
\label{1}
&m_{\SS0}\frac{\DS D\mbox{\frak u}_{\mbox{\sfrak p}}\strut}{\DS d\tau\strut}
-\mbox{\frak S}_{\mbox{\sfrak pq}}
     \frac{\DS D^{\SS2}\mbox{\frak u}^{\mbox{\sfrak q}}\strut}{\DS d\tau^{\SS2} \strut}
=\frac{\DS1 \strut}{\DS 2 \strut}\mbox{\frak u}^{\mbox{\sfrak m}}
     \mbox{\frak R}_{\mbox{\sfrak mpnq}}\mbox{\frak S}^{\mbox{\sfrak nq}}\,,
&\\ \nonumber
&\mbox{\frak u}_{\mbox{\sfrak q}}\mbox{\frak u}^{\mbox{\sfrak q}}=1 \,.
&\end{eqnarray}

If only a {\em free} and {\em planar} motion is going to be considered, the
equation (\ref{1}) splits into the following two ($\mu,\nu=0,1,2$):
\begin{eqnarray}
\label{2}
&m_{\SS0}\dot u_{\mu}-S_{\mu\nu}\ddot u^{\nu}=0\,, &
\\ \label{21}
&\mbox{\frak S}_{{\SS3}\mu}\ddot u^{\mu}=0\,. &
\end{eqnarray}

We shall demonstrate later that the equation (\ref{2}) may be cast into the form of being
the {\em only} third-order relativistic equation admitting a lagrangian description.

Applying to it a kind of the `hamiltonization' prescription of \cite{K1}
then yields an equivalent to the Dixon equations \cite{D} adopted to
the case presently considered here, from which the equation (\ref{2})
follows in turn, provided the Pirani supplementary condition
\begin{equation}
\label{3}
\mbox{\frak u}_{\mbox{\sfrak q}}\mbox{\frak S}^{\mbox{\sfrak pq}}=0
\end{equation}
is in force.

\section{LAGRANGIAN DESCRIPTION}
Necessary and sufficient conditions to the existence of a Lagrange function
for a third-order differential equation
\begin{equation}\label{4}
\A{\,\bkey.\,}{\hbox{\ssb v}}^{{{\prime}}{{\prime}}}{\,+\,}({\hbox{\ssb v}}^{{\prime}}{\!\bkey.\,}{\bpartial}_{\eisf v})\,
\A{\,\bkey.\,}{\hbox{\ssb v}}^{{\prime}}{\,+\,}\B{\,\bkey.\,}{\hbox{\ssb v}}^{{\prime}}{\,+\,}{\hbox{\ssb c}}\,=\,\hbox{\ssb 0}
\end{equation}
were established in \cite{M1}. They are expressed by means of the following
system of partial differential equations in the independent variables
$t$, ${{\sf x}}^a$, and ${{\sf v}}^a$
{\renewcommand{\arraystretch}{1.7}
\begin{equation}\label{5}
\begin{array}{c}
(\romannumeral1) \quad      \partial_{_{_{_{{\hbox{\sssm v}}}}}}{\!}_{[a}{}{{\sf A}}_{bc]}=0
\\ (\romannumeral2) \quad       2\,{{\sf B}}_{[ab]}-3\,{\bf D_{_{\bmat 1}}}{\kern.01667em}{{\sf A}}_{ab}=0
\\ (\romannumeral3) \quad    2\,\partial_{_{_{_{{\hbox{\sssm v}}}}}}{\!}_{[a}{}{{\sf B}}_{b]c}
               -4\,\partial_{_{_{_{{\hbox{\sssm x}}}}}}{\!}_{[a}{}{{\sf A}}_{b]c}
               +{\partial_{_{_{_{{\hbox{\sssm x}}}}}}{\!}_{c}}{\,}{{\sf A}}_{ab}
               +2\,{\bf D_{_{\bmat 1}}}{\kern.01667em}{\partial_{_{_{_{{\hbox{\sssm v}}}}}}{\!}_{c}}{\,}{{\sf A}}_{ab}=0
\\ (\romannumeral4) \quad      {\partial_{_{_{_{{\hbox{\sssm v}}}}}}{\!}_{(a}}{}{{\sf c}}_{b)}
               -{\bf D_{_{\bmat 1}}}{\kern.01667em}{{\sf B}}_{(ab)}=0
\\ (\romannumeral5) \quad      2\,{\partial_{_{_{_{{\hbox{\sssm v}}}}}}{\!}_{c}}{\,}\partial_{_{_{_{{\hbox{\sssm v}}}}}}{\!}_{[a}{}{{\sf c}}_{b]}
           -4\,\partial_{_{_{_{{\hbox{\sssm x}}}}}}{\!}_{[a}{}{{\sf B}}_{b]c}
           +{{\bf D_{_{\bmat 1}}}}^{2}{\,}{\partial_{_{_{_{{\hbox{\sssm v}}}}}}{\!}_{c}}{\,}{{\sf A}}_{ab}
           +6\,{\bf D_{_{\bmat 1}}}{\kern.0334em}\partial_{_{_{_{{\hbox{\sssm x}}}}}}{\!}_{[a}{}{{\sf A}}_{bc]}=0
\\ (\romannumeral6) \quad     4\,\partial_{_{_{_{{\hbox{\sssm x}}}}}}{\!}_{[a}{}{{\sf c}}_{b]}
           -2\,{\bf D_{_{\bmat 1}}}{\kern.0334em}\partial_{_{_{_{{\hbox{\sssm v}}}}}}{\!}_{[a}{}{{\sf c}}_{b]}
           -{{\bf D_{_{\bmat 1}}}}^{3}{\,}{{\sf A}}_{ab}=0\,.
\end{array}
\end{equation}
}

We recall that the skew-symmetric matrix $\A$, the matrix $\B$, and the column vector
${\hbox{\ssb c}}$ depend on the variables $t$, ${\hbox{\ssb x}}$, ${\hbox{\ssb v}}={d{\hbox{\ssb x}}}/dt$.
The differential operator ${\bf D_{_{\bmat 1}}}$ denotes the first order generator of the Cartan distribution,
\[
{\bf D_{_{\bmat 1}}}=\partial_{t}{\,+\,}{\hbox{\ssb v}}{\,\bkey.\,}{\bpartial}_{\eisf x}\,.
\]
Any third-order Euler-Poisson equation (i.e. that of some variational origin)
in two space dimensions fits into the form (\ref{4}).

We are interested only in the equations bearing the Poincar\'e symmetry with
the infinitesimal generator $X$ parametrized by means of a skew-symmetric
matrix ${\bf\Omega}$ and some vector
\boldmath
$\pi$:
\b
\eqnarray
\ubm X&=&
-(\pi\cdot\hbox{\ssb x})\,\ubm{\partial_t}+\ubm{g_{\SS00}\,t}\,\pi\,.\,\partial_{\eisf x}
+M\cdot(\hbox{\ssb x}"\partial_{\eisf x})
\nonumber\\ &&{}+\ubm{g_{\SS00}}\,\pi\,.\,\partial_{v}
+(\pi\cdot V)\,V\,.\,\partial_{v}+M\cdot(V"\partial_{v})\nonumber
\\ \nonumber &&{}+\ubm2\,(\pi\cdot V)\,W.\,\partial{_{w}}+(\pi\cdot W)\,V\,.\,\partial{_{w}}+M\cdot(W"\partial{_{w}})\,\ubm.
\endeqnarray
\e

The centered dot symbol denotes the inner product of vectors or tensors;
the lowered dot symbol denotes
the contraction of a row-vector and the subsequent column-vector,
or (sometimes) the contraction of a matrix and the subsequent column-vector.
In order to chose a convenient expression of the symmetry concept we introduce
a vector differential form $\epsilon$, associated with the equation (\ref{4}):
\begin{eqnarray}
\ubm\epsilon_{a}={\sf A}_{ab}{\rm d}{\sf v}^{\prime}{}^{b}&{}+{}&{\sf k}_{a}{\rm d}\ubm t\,,\nonumber
\\ &&\b\hbox{\ssb k}=(W\cdot\partial_{v})\,AW+\BW+\c\e\,\ubm.\lable{6}
\end{eqnarray}

Now it is possible to cast the idea of the symmetry of the equation (\ref{4})
into the framework of the concept of exterior differential system invariance.
The system in case is generated by the vector valued Phaff  form $\epsilon$ and
the contact vector valued differential forms
\begin{equation}
\label{61}
\b {\rm d}\hbox{\ssb x}-V{\rm d}\ubm t\,,\qquad {\rm d}V-W{\rm d}\ubm t\,\e\ubm.
\end{equation}
Let $\ubm X(\epsilon)$ denote the Lie derivative of the vector valued differential form
$\epsilon$ along the vector field $\ubm X$. The invariance condition consists in that
there may be found some matrices ${\bf\Phi}$, ${\bf\Xi}$, and ${\bf\Pi}$
depending on $\copy\Elocity$ and $\copy\PrimE$ such that
\begin{equation}\lable{7}
\b X(\epsilon)={\bf\Phi}\,.\,\epsilon+{\bf\Xi}\,.\,({\rm d}\hbox{\ssb x}-V{\rm d}\ubm t)
+{\bf\Pi}\,.\,({\rm d}V-W{\rm d}\ubm t)\e\ubm.
\end{equation}
We also assert that $\A$ and $\hbox{\ssb k}$ in (\ref{6}) do not depend neither on $\ubm t$ nor on $\hbox{\ssb x}$.

The identity (\ref{7}) splits into more identities, obtained by evaluating the
coefficients of the differentials ${\rm d}\ubm t$, ${\rm d}\hbox{\ssb x}$,
$\b{\rm d}V\e$, and $\b{\rm d}W\e$ independently:
\b
\begin{eqnarray}
\nonumber
&{\big(\pi\,.\,\partial_{v}+(\pi\cdot V)\,V\,.\,\partial_{v}+M\cdot(V"\partial_{v})\big)\,A}
\hspace{2.9cm}
&\\ \lable{81}&{{}+\ubm2\,(\pi\cdot V)A+(AV)\otimes\pi-AM={\bf\Phi}A}\,;&
\\ \lable{82}&\ubm2\,(AW)\otimes\pi+(\pi\cdot W)\,A\,={\bf\Pi}\,;&
\\ \lable{83}&-\hbox{\ssb k}\otimes\pi={\bf\Xi}\,;&
\\ \lable{84}&\ubm X(\hbox{\ssb k})={\bf\Phi}\hbox{\ssb k}-{\bf\Xi}V-{\bf\Pi}W\,\ubm.&
\end{eqnarray}
\e
In the above the `$\otimes$' symbol means the tensor (sometimes named as
`direct') product of matrices.

A skew-symmetric two-by-two matrix always has the inverse, so the
`Lagrange multipliers' ${\bf\Phi}$, ${\bf\Xi}$, and ${\bf\Pi}$ may explicitly be
defined from the equations (\ref{81}--\ref{83}) and then substituted into~(\ref{84}).
Subsequently, the equation (\ref{84}) splits into the following identities
by the powers of the variable $\copy\PrimE$ and by the parameters ${\bf\Omega}$
and $\pi$ (take notice of the derivative matrix $\b A^{\prime}=(W.\,\partial_{v})\,A\e$;
also the vertical arrow sign points to the very last factor to which the
aforegoing differential operator still applies):
\b
\begin{eqnarray}
\nonumber
     \lefteqn{\bigl(M\cdot(V"\partial_{v})\bigr)\,A^{\prime}W+\bigl(M\cdot(W"\partial_{v})\bigr)\,AW
     -(W.\,\partial_{v})\,AMW}
\\   \lable{95}
         &{{}=\bigl(M\cdot(V"\partial_{v})\bigr)\,\sa A^{-1}A^{\prime}W-AMA^{-1}A^{\prime}W}\,;
&\\ &\lable{96}\stack\bigl(M\cdot(V"\partial_{v})\bigr)\,\B-\BM
     =\bigl(M\cdot(V"\partial_{v})\bigr)\,\sa A^{-1}\B-AMA^{-1}\B\,;
&\\ &\lable{97}\stack\bigl(M\cdot(V"\partial_{v})\bigr)\,\c
     =\bigl(M\cdot(V"\partial_{v})\bigr)\,\sa A^{-1}\c-AMA^{-1}\c\,;
&\\ \nonumber
     \lefteqn{\stack\bigl(\pi\,.\,\partial_{v}+(\pi\cdot V)\,V\,.\,\partial_{v}\bigr)\,A^{\prime}W
     +(\pi\cdot V)\,A^{\prime}W+(\pi\cdot W)\,(V\,.\,\partial_{v})\,AW+(\pi\cdot W)\,A^{\prime}V}
\\ &   \label{98}{}=\bigl(\pi\,.\,\partial_{v}+(\pi\cdot V)\,V\,.\,\partial_{v}\bigr)\,\sa A^{-1}A^{\prime}W
          +(\pi A^{-1}A^{\prime}W)\,AV-\ubm3 \,(\pi\cdot W)\,AW\,;
&\\ \nonumber
     \lefteqn{\stack\bigl(\pi\,.\,\partial_{v}
     +(\pi\cdot V)\,V.\,\partial_{v}\bigr)\,\B+(\BV)\otimes\pi}
\\ &\label{99}
        {}=\bigl(\pi.\,\partial_{v}+(\pi\cdot V)\,V\,.\,\partial_{v}\bigr)\,\sa A^{-1}\B
        +(AV)\otimes\pi A^{-1}\B+(\pi\cdot V)\B\,;
&\\ \nonumber
     \lefteqn{\stack\bigl(\pi\,.\,\partial_{v}+(\pi\cdot V)\,V\,.\,\partial_{v}\bigr)\,\c}
\\ &\label{90}{}=\bigl(\pi\,.\,\partial_{v}+(\pi\cdot V)\,V\,.\,\partial_{v}\bigr)\,\sa A^{-1}\c
     +\ubm3 \,(\pi\cdot V)\,\c+(\pi A^{-1}\c)\,AV\,.
\hspace{-1cm}&\hspace{1cm}\end{eqnarray}
\e
\unboldmath

Cumbersome although routine calculations accompanying the simultaneous
solving of the partial differential equations (\ref{95}) and (\ref{98})
with respect to the unknown function ${\sf A}_{12}$ produce the unique
output of
\[
{\sf A}_{12}=\frac{\rm const}{(1+{\sf v}_{\SS1}{\sf v}^{\SS1}+{\sf v}_{\SS2}{\sf v}^{\SS2})^{3/2}\strut}\,.
\]

We remind that the system of the equations (\ref{95}--\ref{90}) and the
system~(\ref{5}) must be solved simultaneously. Thus, the
equation (\ref{5}\romannumeral1) becomes trivial now.

Under the assumption of $\B$ being a symmetric matrix (see (\ref{5}\romannumeral2)),
the solution of the equations \{(\ref{96}), (\ref{99})\} amounts to
\[\b
{\sf B}_{ab}={\rm const}\cdot(1+V\bcdot V)^{-3/2}\bigl[\e{\sf v}_{a}{\sf
v}_{b}\b-(1+V\bcdot V)\,{\sf g}_{ab}\bigr]\,. \e
\]

This automatically satisfies  the equation (\ref{5}\romannumeral3) too.
In what concerns the subsystem \{(\ref{97}), (\ref{90})\}, only the trivial solution
$\c={\bs0}$ exists.

We are ready now to formulate the summary of the above  development in terms of a proposition:
\begin{prop}
The invariant Euler-Poisson equation of a relativistic planar motion is:
\begin{equation}
\label{10}\b
-\frac{\astV''\strut}{(1+V\bcdotV)^{3/2}\strut}+3\,\frac{\astW\strut}{(1+V\bcdotV)^{5/2}}\,(V\bcdotW)
+\frac{\mu\strut}{(1+V\bcdotV)^{3/2}}\,\bigl[(1+V\bcdotV)\,W-(W\bcdotV)\,V\bigr]={\bs0}\,.
\e
\end{equation}
\end{prop}

Arbitrary constant $\mu$ serves to parametrize the set of all the variational equations (\ref{10}).
The definition of the `star operator' is common. Thus, $\ast1=\bs e_{(\SS1)}\wedge\bs e_{(\SS2)}$
whereas $\ast\,(\bs e_{(\SS1)}\wedge\bs e_{(\SS2)})=1$ if the (pseudo)\-\kern.5pt orthonormal
frame $\{\bs e_{(\SS1)}\,,\ \bs e_{(\SS2)}\}$ carries the positive orientation;
also $(\ast\mbox{\ssb w})_{a}=\varepsilon_{ba}{\sf w}^{b}$. We recall for future use
the definition of the inner product of two bi-vectors:
\[
(\bs a\wedge\bs b)\bcdot(\bs c\wedge\bs d)\,
=\,(\bs a\bcdot\bs c)\,-\,(\bs b\bcdot\bs d)\,.
\]

\begin{prop}
The Euler-Poisson equation (\ref{10}) describes the free motion of a spinning
particle in two space dimensions if
\begin{equation}
\label{101}
u_{\nu}S^{\mu\nu}=0\,.
\end{equation}
\end{prop}

{\em Demonstration.} Equation (\ref{10}) describes the world line of a particle parametrized
by time. Passing to the proper time parametrization one obtains:
\boldmath
\begin{equation}
\label{11}
{\bf\ddot{\bmat u}}\times u+\ubm \mu {\bf\dot{\bmat u}}= {\bf0}\,\ubm.
\end{equation}
\unboldmath

Let us introduce a vector $a_{\mu}=\frac{1}{2}\varepsilon_{\nu\lambda\mu}S^{\nu\lambda}$.
Then  the equation (\ref{2}) takes the form
\boldmath
\begin{equation}
\label{12}
\ubm{m_{\SS0}} {\bf\dot{\bmat u}}+a\times{\bf\ddot{\bmat u}}={\bf0}
\end{equation}
with the consequence that
\begin{equation}
\label{13}
a\cdot{\bf\dot{\bmat u}}={\bf0}\,\ubm.
\end{equation}
Any vector $a$ may always be presented as
\begin{equation}
\label{131}
a=(a\cdot u)\,u-(a\times u)\times u\,\ubm.
\end{equation}
If the solutions of the equation (\ref{11}) are to satisfy also the equation (\ref{12}),
we may define the variable ${\bf\ddot{\bmat u}}$ from (\ref{11}) with the
help of
$\;({\bmat u}\cdot{\bf\ddot{\bmat u}})
=-({\bf\dot{\bmat u}}\cdot{\bf\dot{\bmat u}})\;$
and substitute it into
(\ref{12}) to obtain, in view of (\ref{13}),
\begin{equation}
\label{14}
-\ubm{\mu}\,(a\cdot u)\,{\bf\dot{\bmat u}}-\ubm{m_{\SS0}}{\bf\dot{\bmat
u}}+({\bf\dot{\bmat u}}\cdot{\bf\dot{\bmat u}})\,a\times u={\bf0}\,\ubm.
\end{equation}
The condition (\ref{101}) is equivalent to $\,a\times u=0$, thus the
equation (\ref{14}) gives
\begin{equation}
\label{15}
(a\cdot u)=-\ubm{m_{\SS0}/\mu}\,\ubm,
\end{equation}
\unboldmath
and from (\ref{131}) it follows that
\begin{equation}
\label{16}
\bmat a=-\frac{m_{\SS0}}{\mu}\,\bmat u\,,
\end{equation}
so the equations (\ref{11}) and (\ref{12}) are now equivalent.
\smallskip

{\em Remark~1.} In most general setting,
when $\|\br u\|\not=1$ and $|\fr g|\not=1$ ($\fr g=\det(\fr g_{\sr{mn}})$,
$\fr m$, $\fr n$ run from 0 to 3) the four-vector of spin is introduced by
\begin{equation}
\label{17}
\fr s_{\mbox{\sfrak p}}=\frac{\sqrt{|\fr g|}}{2\|\mbox{\br u}\|}\,\varepsilon_{\mbox{\sfrak mnqp}}
     \mbox{\frak u}^{\mbox{\sfrak m}}\mbox{\frak S}^{\mbox{\sfrak nq}}\,.
\end{equation}
It is straightforward that $\fr s_{\SS3}=-a_{\mu}u^{\mu}$ and thus one gets a
`renormalization' of the spinning particle's mass:
\begin{equation}
\label{171}
\mu=\frac{m_{\SS0}}{\fr s_{\SS3}}\,.
\end{equation}

{\em Remark~2.} The first space-time curvature of the particle's world line
governed by (\ref{11}) is constant, i.e. $\frac{\DS\strut d}{\DS\strut d\tau}\|{\bf\dot{\bmat u}}\|=0$.

{\em Remark~3.} The point symmetries of (\ref{10}) are being exhausted by
pseu\-do-or\-tho\-go\-n\-al (resp. conformal) transformations if $m\ne0$
(resp.~$m=0$). The proof may be found in \cite{9A}.
\smallskip

We can present two different
($a=1,2$) Lagrange functions which produce the equation (\ref{10}),
\begin{equation}
\label{18}
\b
L_{(a)}=\frac{\ast(V^{\prime}"\bs e_{(a)})}{(1+V\bcdotV)^{1/2}(1+{\sf g}_{aa}\|V"\bs e_{(a)}\|^{2})}
     \e{\sf v}^{a}-\mu\,\b (1+V\bcdotV)^{1/2}\,.\e
\end{equation}
These differ by the total time derivative:
\[
L_{(2)}-L_{(1)}=\frac{d}{dt}\arctan\frac{{\sf v}^{\SS1}{\sf v}^{\SS2}}{\sqrt{1+{\sf v}_{a}{\sf v}^{a}}}\,.
\]

Now it's time to pass over to the hamiltonian counterpart exposition.
\section{HAMILTONIAN DESCRIPTION}
A detailed exposition of the (generalized) hamiltonian theory applicable to the
odd-order differential equations at no less extent than to the even-order ones may be found
in \cite{K1}; a concise exposition is presented in \cite{K2}.

A conjectural Legendre transformation is given by the momenta
\newbox\pprime\setbox\pprime=\hbox{\bs p$^{\prime}$}
\newbox\pbox\setbox\pbox=\hbox{\bs p}\wd\pprime=\wd\pbox
\arraycolsep=\colsep
\boldmath\b
\begin{eqnarray*}
\bs p&=&\frac{\partial\ubm L}{\partialV}
-\frac{\ubm d}{\ubm{dt}}\,\frac{\partial\ubm L}{\partialW} \\[2mm]
\box\pprime&=&\frac{\partial\ubm L}{\partialW}\,\ubm.
\end{eqnarray*}
\e\unboldmath
The Hamilton function
\[H=-L+\bs p\bkey.\bs v+\bs p^{\prime}\!\bkey.\bs v^{\prime}\]
in case of a regular Legendre transformation gives rise to the following Hamilton
system of first order equations:
\begin{eqnarray}
\label{19}
-\frac{\partial H}{\partial{\sf x}^a}-\frac{\partial{\sf v}^b}{\partial{\sf x}^a}
\frac{d}{dt}{\sf p}^{\prime}_b-\frac{d}{dt}{\sf p}_{a}&=&0\,; \\[2mm]
\label{190}
-\frac{\partial H}{\partial{\sf p}_{a}}+\frac{d}{dt}{\sf x}^{a}&=&0\,;\\[2mm]
\label{192}
-\frac{\partial H}{\partial {\sf p}^{\prime}_{a}}+\frac{\partial{\sf v}^{a}}{\partial{\sf x}^{b}}
     \frac{d}{dt}{\sf x}^{b}+\left(\frac{\partial{\sf v}^{a}}{\partial{\sf p}^{\prime}_{b}}
     -\frac{\partial{\sf v}^{b}}{\partial{\sf p}^{\prime}_{a}}\right)\frac{d}{dt}{\sf p}^{\prime}_{b}&=&0\,.
\end{eqnarray}

Both $L_{\SS1}$ and $L_{\SS2}$ from (\ref{18}) produce one and the same Hamilton
function:
\begin{eqnarray}
\nonumber
H&=&\frac{\ast(\bs v\wedge \bs v^{\prime})}{(1+\bs v\bcdot\bs v)^{3/2}}+\mu(1+\bs v\bcdot\bs v)^{-1/2}\\[2mm]
\label{191}
 &=&\bs p\bkey.\bs v+\mu\sqrt{1+\bs v\bcdot\bs v}\,.
\end{eqnarray}
This happens because both $L_{\SS1}$ and $L_{\SS2}$ lead to the same `zero-order'
momentum,
\begin{equation}
\label{20}
\bs p=\b\frac{\astW}{(1+V\bcdot V)^{3/2}}-\mu\,\frac{V}{\sqrt{1+V\bcdot V}}\e\,.
\end{equation}
However, the `first-order' momentum $\bs p^{\prime}$ has the identically equal to
zero first (resp. second) component if one starts with $L_{\SS1}$ (resp. $L_{\SS2}$)
alone. This suggests that we take
\[L=\frac{1}{2}\bigl(L_{\SS1}+L_{\SS2}\bigr)\]
in an attempt to proceed further with a kind of regular Legendre transformation.
In this case one calculates out the following expressions for the two components
of the momentum $\bs p^{\prime}$:
\begin{equation}
\label{201}
{\sf p}'_{1}=\frac{{\sf v}^{2}}{2\sqrt{1+\bs v\bcdot\bs v}\,(1+{\sf v}_{1}{\sf v}^{1})}\,,\quad
{\sf p}'_{2}=-\,\frac{{\sf v}^{1}}{2\sqrt{1+\bs v\bcdot\bs v}\,(1+{\sf v}_{2}{\sf v}^{2})}\,.
\end{equation}

In (\ref{192}) the inverse to the Jacobi matrix of the Legendre transformation is indispensable
to know
(at least in its $\frac{\DS\strut\partial\bs v}{\DS\strut\partial\bs p'}$ part).
The Jacobi matrix itself is readily obtained from (\ref{20}) and (\ref{201}),
so we cite below only that part of its inverse, important for the
forthcoming calculations,
{
\def\v#1{{\sf v}_{#1}}\def\V#1{{\sf v}^{#1}}
\begin{eqnarray}
\nonumber
\null&\left(\frac{\DS\partial{\sf v}^{a}\strut}{\DS{\,}\partial{\sf p}'{}_{b}\strut}\right)=
     \frac{\DS2\strut}{\DS\Delta\strut}(1+\b V\bcdot V \e)^{3/2}\times&\null \\[3mm]
\label{22}
\null&{}\times\left(
\begin{array}{cc}
\frac{\DS\strut\v2\V1(3+3\,\v2\V2+2\,\v1\V1)}{\DS\strut(1+\v2\V2)^{2 \strut}}&-1
\\ 1&-\frac{\DS\strut\v1\V2(3+3\,\v1\V1+2\,\v2\V2)}{\DS\strut(1+\v1\V1)^{2 \strut}}
\end{array}
\right)&\null\,,
\end{eqnarray}
}
where $\Delta$ denotes the determinant of the matrix in (\ref{22}).
The Reader may also easily convince himself by direct calculation that
\begin{equation}
\label{23}
\left(\frac{\partial{\sf v}^{a}}{{\;}\partial{\sf p}{\;}_{b}}\right)=\bs0\,.
\end{equation}

As far as the Hamilton function (\ref{191}) does not depend on the space and time
variables, the essential part of the Hamilton system, the equation (\ref{19}), constitutes nothing
else but the conservation of the momentum,
\begin{equation}
\label{24}
\frac{d\bs p}{dt}=\bs0\,.
\end{equation}
The remaining two equations mean merely that we pick up only the holonomic
solutions of (\ref{24}). Thus, the equation (\ref{190}) in view of (\ref{23})
reads
\[-\bs v+\frac{d\bs x}{dt}=\bs0\,,\]
and so it describes the kernel of the first one of the two differential forms
in (\ref{61}). The equation (\ref{192}) in view of (\ref{22}) reads
\begin{equation}
\label{26}
-\frac{\bpartial H}{\bpartial\bs p'}+4\frac{(1+\bs v\bcdot\bs v)^{3/2}}{\Delta}\,
\left(\ast\,\frac{d\bs p'}{dt}\right)=\bs0\,.
\end{equation}
But we can calculate $\frac{\DS\strut\bpartial H}{\DS\strut\bpartial\bs p'}$ from (\ref{191})
by means of (\ref{22}) and also differentiate the definitions (\ref{201})
explicitly. After substituting into (\ref{26}) it becomes evident that the
Hamilton equation (\ref{192}) describes the kernel of the second
differential form in (\ref{61}). In fact the equation (\ref{26}) transforms
into the following system: \[\bs M\,\bkey.\,(d\bs v-\bs v'dt)=\bs0\,,\]
where $\bs M$ is the matrix from (\ref{22}).

\medskip

Let us return to the gravitational dipole particle.
The Dixon system of equations in the general curved four-dimensional space-time
accepts arbitrary parametrization of the particle's world line. It reads:
\begin{eqnarray}
\label{27}
\frac{d\fr P}{d\lambda}&=&\fr F \\[2mm]
\label{28}
\frac{d\br S}{d\lambda}&=&2\,\br P\wedge\br u\,.
\end{eqnarray}
The force $\br F$ depends, apart from space and time variables, also upon the
particle's tensorial spin $\br S$ and velocity $\br u$; it must satisfy
the constraint $\,\br F\bkey.\,\br u=\br0\,$. Once supplemented by the
Pirani constraint (\ref{3}), the system (\ref{27}, \ref{28}) allows an
equivalent transcription in terms of the spin four-vector (\ref{17}),
`resolved' with respect to the variable $\br P$. Namely, in place of
(\ref{28}) we can write (renouncing the constraint $\|\br u\|=1$)
\begin{eqnarray}
\label{29}
&\br P=\frac{\DS\strut m_{\SS0}}{\DS\strut\|\br u\|}\,\br u-({\rm sgn}\,\fr g)\,\ast\,({\bf\dot{\br u}}\wedge\br u\wedge\br s)\,,&\\[2mm]
\nonumber&\bf{\dot{\br s}}\wedge\br u=\br0\,.&
\end{eqnarray}
In the parametrization by time ($\fr u^{\SS0}=1$), denoting by ${\bf P}$ the three-vector
part of the four-vector $\br P$, the definition (\ref{29}) takes up the shape
\begin{equation}
\label{30}
{\bf P}=m_{\SS0}\,\frac{\bf v}{\sqrt{1+{\bf v}\bcdot{\bf v}}}+\frac{{\rm sgn}\,\fr g}{(1+{\bf v}\bcdot{\bf v})^{3/2}}
     ({\bf v}'\times{\bf s}-\fr s^{\SS0}\,{\bf v}'\times{\bf v})\,,
\end{equation}
where we use the notation ${\bf v}$ for the three-vector part of the velocity
four-vector $\br u$ in the special case when $\fr u^{\SS0}=1$; also $\br s=(\fr s^{\SS0}, {\bf s})$
according to the general template.

Reassuming the particle's motion be two-dimensional only, one obtains from (\ref{30})
\begin{equation}
\label{31}
\bs P=\frac{m_{\SS0}\bs v}{\sqrt{1+\bs v\bcdot\bs v}}-({\rm sgn}\,\br g)\,\fr s^{\SS3}\,
     \frac{\ast\bs v'}{(1+\bs v\bcdot\bs v)^{3/2}}\,.
\end{equation}
It suffices to compare (\ref{31}) with the expression (\ref{20}),
assuming the notation (\ref{171}), to come
immediately to the conclusion that
the quantity ${\bs P}$ coincides with $\fr s^{\SS3}$ times the canonical hamiltonian
momentum.

\begin{prop}
The Hamilton function for the planar motion of free relativistic spinning particle
with the mass (\ref{171}) is given by (\ref{191}).
\end{prop}

{\sl Remark~4}. The Mathisson equation in general-relativistic framework and
under arbitrary world line parametrization was considered in \cite{M2}.
A set of the Lagrange functions to produce it in flat space-time was
suggested in \cite{M3}.
A second-order differential-geometric connection related to the equation
(\ref{10}) was constructed in \cite{M4}.

\label{last@page}
\end{document}